\documentclass[12pt]{article}
\usepackage{epsfig}
\usepackage{rotating}
\begin{document}

\thispagestyle{empty}

\begin{center}
{\Large{\bfseries The ngdp framework
for data acquisition systems
}}\\[10mm]
{\Large{\bfseries A.Yu.~Isupov}}\\[10mm]
{\itshape Veksler and Baldin Laboratory of High Energy Physics}\\[5mm]
{\itshape Joint Institute for Nuclear Research}
\end{center}


\newpage

\centerline{\bfseries Abstract}

\vspace*{5mm}

\noindent Isupov~A.Yu. \\
The ngdp framework
for data acquisition systems

\vspace*{5mm}

The {\itshape ngdp} framework is intended to provide a base for the
data acquisition (DAQ) system software.
The {\itshape ngdp}'s design key features are:
high modularity and scalability;
usage of the kernel context (particularly kernel threads)
of the operating systems (OS), which allows to avoid
preemptive scheduling and unnecessary
memory--to--memory copying between contexts;
elimination of intermediate data storages on the media slower than the
operating memory like hard disks, etc.
The {\itshape ngdp}, having the above properties, is suitable to organize
and manage data
transportation and processing for needs of essentially distributed
 DAQ systems.

The investigation has been performed at the Veksler and Baldin Laboratory
of High Energy Physics, JINR.

\vspace*{1cm}

\centerline{\bfseries Аннотация}

\vspace*{5mm}

\noindent Исупов~А.Ю. \\
Инфраструктурная система ngdp как основа для реализации систем сбора данных

\vspace*{5mm}

Описана инфраструктурная система (framework) {\itshape ngdp}, предоставляющая
основу для создания программного обеспечения систем сбора данных (DAQ).
Ключевые особенности {\itshape ngdp} следующие:
хорошие модульность и масштабируемость;
реализация основной части кода в контексте ядра операционной системы (в
частности, в виде нитей ядра), позволяющая избежать принудительного
планирования и излишних копирований памяти между контекстами;
исключение промежуточных размещений данных на более медленных, чем
оперативная память, носителях (жесткие диски и т.п.).
Таким образом, {\itshape ngdp} пригодна для организации и управления
передачей и обработкой данных в существенно распределенных
 системах DAQ.

Работа выполнена в Лаборатории физики высоких энергий им. В.И.Векслера и
А.М.Балдина ОИЯИ.

\newpage



\setcounter{page}{1}

\section{Introduction}
\label{ngdp.intro}

\vspace*{-2mm}

\hspace*{4mm} Modern experimental physic setups can produce extremely large
data volumes very quickly -- faster, than it can be transferred through single
10~Gbits/sec Ethernet link, so they require more than one link used in
parallel. It means that such setups should be equipped with essentially
distributed (between many computers) data acquisition (DAQ) systems.
On the other hand, the whole dataset belonging to
some physical event should appear at some stage on a single computer for full
event building. This requirement is necessary
for each event. Consequently, this
system should contain more than one Event Builders (EvB) in principle.
This fact requires to solve tasks related
with data streams organization and management:\\
$\bullet$ to merge different data of data flows;\\
$\bullet$ to split the identical streams or duplicate them at another stages;\\
$\bullet$ to provide intermediate buffers and delays, etc.\\
Software for this DAQ system should contain some kind of the data
transportation and processing system able to organize and manage these
data flows. The system should
provide maximal performance and throughput practically reachable
on the generic computer and network hardware, at least, faster than
1~Gbyte/sec. For the software system it means that it
should be as lightweight and fast
as possible: it uses the corresponding design not to consume an essential
resource fraction for execution of its own code.
For the used operating system (OS) it means that the network service itself
should not consume an essential resource fraction either, for example,
execution of TCP/IP stack and Ethernet interface interrupt handlers, etc.
We should localize overall systems ``bottleneck'' in the network as the
slowest system's element to preserve the major fraction of
computer resources for needs of the data processing itself. The
above requirements for the software side can be achieved by the following
means:\\[-8mm]
\begin{itemize}
\item elimination of intermediate storage on slow media (hard disks, etc.);\\[-8mm]
\item minimization of memory--to--memory copying where possible (in
particular, elimination of copying between the user and kernel contexts);\\[-8mm]
\item execution in the almost real--time mode (by means of kernel context
implementation based on the kernel threads);\\[-8mm]
\item the data in the user context should be presented in the form of streams
or memory objects (but not files).\\[-8mm]
\end{itemize}
The proposed system should be
reasonably modular,
easy in implementation, maintenance and usage,
based as much as possible on the existing freely distributable software
packages and technologies.

Through the presented text the references to terms are
highlighted as
{\bfseries boldface text}, file and software package names -- as
{\itshape italic text}, C and other languages constructions -- as
\verb|typewriter text|. Reference to the manual page named ``qwerty'' in the
9$^{\mbox{\small th}}$ section is printed as {\bfseries\itshape qwerty(9)},
reference to the sections in this paper --
as ``section \ref{ngdp.des_impl.transp.fifo}''. Note also verbal
constructions like ``{\bfseries\itshape accept(2)}ed'' and ``\verb|rmhook|ing'',
which means ``accepted by {\bfseries\itshape accept(2)}'' and
``hook removing by \verb|rmhook|''.
Subjects of substitution by actual values are
enclosed in the angle brackets: \verb|<num_of_packets>|, while some optional
elements are given in the square brackets: $[$~\verb|ng_filter|~$\to ]$.
All the mentioned trademarks are properties of their respective owners.

\vspace*{-5mm}

\section{Overview}
\label{ngdp.overview}

\vspace*{-2mm}

\hspace*{4mm} First of all, we should choose a computing environment:
hardware architecture, OS, programming language(s) and
corresponding instrumental software, -- to
design, implement, maintain and use our DAQ software.

On the one hand, we have no special requirements to computers hardware --
other than performance. On the other hand, a big DAQ system can require from
tens to some hundreds
of units of such hardware with corresponding maintenance, etc.
So, we should choose the most standard and generic hardware reasonably cheap
due to a great volume of production.
This architecture called {\ttfamily AMD64/EM64T}, previously
known also as {\ttfamily x86-64} and {\ttfamily IA-32e}, should be used
currently and in the near future.

The operating system used on the online computer determines the DAQ system
design and organization, consequently the inadequate OS selection are sure to
strongly complicate implementation, maintenance,
and using of the DAQ system. The OS itself should have adequate technical
abilities for easy multiple installations, remote maintenance and backup,
read--only boot filesystem and diskless boot, boot without input and
output devices, etc.

UNIX--like OSs are optimal for the above requirements.
UNIX is a multiprocess and multiuser OS with powerful mechanisms for interprocess
and inter--computer communications, a very advanced virtual memory subsystem,
support of sophisticated networking and graphics interfaces, extended tools
for the software design. Costs for UNIX working itself are rather modest and
negligible. Free sources distributions availability of
UNIX--like OSs is a mandatory requirement in our case. After all, high
portability of UNIX programming and approximately
unlimited quantity of the existing software are also very attractive.

To achieve the reasonable performance, we should choose C programming language
(or C++ -- only in such cases, where we can't avoid an object--oriented design
and implement it on C) and ultimately
avoid interpreted languages like Perl or CINT.

Lets briefly remind the basic principles used by {\itshape qdpb} framework
\cite{IsupJINRC01-116} which are still important for the presented design,
too:\\
\noindent $\bullet$
Distributed (between CPUs and computers) DAQ system
is unavoidably split into software modules
interconnected with experimental data streams.\\
$\bullet$
A modular design allows one to separate code pieces dependent of the
experimental setup hardware, experimental data contents and
layout from other ``invariant'' modules%
.\\
$\bullet$
``Invariant'' modules are grouped into some universal
framework suitable for using again and again during construction and upgrade
of DAQ systems. ``Invariant'' modules are intended mostly for data streams
management.\\
$\bullet$
Experimental data are represented in the unified form by packets
(sequences of bytes) contain
the packet header followed by the packet body%
:\\
--
Packet header\index{{Packet}!{header}} has a fixed size
and format and contains at least the following fields:
packet identifier, packet length, packet type, packet
serial number, packet creation time and packet check sum (CRC).
The packet identifier is identical for all packets.
Packets of different types have separate serial numeration%
.\\
--
Packet body is experimental data of a single event (trigger) itself,
encapsulated into the packet for transportation purposes, and has the
known length.
The packet length is not coupled with the packet type -- in other words,
the bodies with different length are permitted for the same packet type.
The packet size is limited by the \verb|PACK_MAX| value.
Additionally to data packets the control packets and packets of response to
control packets (the so called ``answer packets'')
should be implemented, too.\\
$\bullet$
Streams of such packets can:\\
--
be transferred locally (on single computer) and/or\\
--
remotely (between different computers through network);\\
--
cross the context boundaries from the kernel space to the
user one and vice versa;\\
--
be buffered, copied, filtered, merged in a different manner, etc.\\
Note, all these activities are carried out exclusively in the memory.
Intermediate
storages on slow media like hard disks (HDD) are eliminated.\\
$\bullet$
Software modules can be implemented as processes in the user context
and as the so called loadable kernel modules (KLD) -- in the kernel
context.\\
$\bullet$
Packet streams between processes are implemented by unnamed pipes
locally and by socket pairs -- remotely.

However, more than ten years of computing technologies progress after
early {\itshape qdpb} variants implementing, has allowed us to use the
following in
our design:\\
$\bullet$
Modern kernels allow to execute some code
pieces in the kernel context -- the so called ``kernel threads'' -- autonomously
like processes in the user context in contrast with
traditional kernels, whose code can be executed only in the result of
external events: system call by process, interrupt request (IRQ), etc. Note,
such threads are not subjects for preemptive scheduling and voluntarily release
CPU. Due to the kernel threads we can fulfil most of the packet processing as
fast as possible and in the same kernel context where the packets originate
from hardware drivers or network sockets.\\
$\bullet$
So, we need tools for packet stream management within the kernel.
Fortunately, these tools already exist, and one of them is
the {\bfseries\itshape netgraph(4)} package, after which our framework is named
{\itshape ngdp} -- \underline{n}et\underline{g}raph based \underline{d}ata
\underline{p}rocessing. Originally {\bfseries\itshape netgraph(4)} was used
to distribute network packets between some nodes to implement the network
protocol layers. Lets cite from the corresponding manual pages:
``The netgraph system provides a uniform and modular system for the
implementation of kernel objects which perform various networking functions.
The objects, known as {\bfseries nodes}, can be arranged into
arbitrarily complicated
graphs. Nodes have {\bfseries hooks} which are used to connect
two nodes together,
forming the edges in the graph. Nodes communicate along the edges to
process data, implement protocols, etc... All nodes
implement a number of predefined methods which allow them to interact
with other nodes in a well defined manner.
Each node has a type, which is a static property of the node determined
at node creation time.''
In the {\bfseries\itshape netgraph(4)} the data are flowing
along the graph edges while control messages are delivered directly from
the source to destination.\\
$\bullet$
From the object oriented programming (OOP)
point of view, the node types are classes, nodes are instances of their
respective class, and interactions between them are carried out via well
defined interfaces.\\
$\bullet$
The modular design of the proposed basic framework allows us to easy maintain
the essentially distributed software system due to high scalability of the
{\bfseries\itshape netgraph(4)}. On each computer we can produce an arbitrary
number of instances of some node type limited only by the available memory.

The {\bfseries\itshape netgraph(4)} package provides
the following entities of our interest:\\[-8mm]
\begin{itemize}
\item
 socket {\bfseries\itshape ng\_ksocket(4)} for the remote
data transfer by IP protocol (TCP or UDP);\\[-8mm]
\item
 socket {\bfseries\itshape ng\_socket(4)} for data and control messages
interchange between the kernel context graph and the user context process;\\[-8mm]
\item
 {\bfseries\itshape netgraph(3)}
library to simplify control over {\bfseries\itshape ng\_socket(4)} and
transfer through it for the user context processes;\\[-8mm]
\item
 means for building the graph
itself: infrastructure in the kernel --
\verb|netgraph| KLD
module, -- and
{\bfseries\itshape ngctl(8)}, {\bfseries\itshape nghook(8)}
utilities;\\[-8mm]
\item
 service nodes for data flow managing:
{\bfseries\itshape ng\_tee(4)}, {\bfseries\itshape ng\_one2many(4)},
{\bfseries\itshape ng\_split(4)}%
;\\[-8mm]
\item
 nodes for debugging: {\bfseries\itshape ng\_source(4)},
{\bfseries\itshape ng\_hole(4)}, {\bfseries\itshape ng\_echo(4)}.\\[-8mm]
\end{itemize}

Lets assume that a big DAQ system will split into
logical levels of data processing along the data flow%
as follows:\\
$\bullet$ FEM
(Front--End Modules)
level -- standalone computers and/or processor modules in crates of
read--out electronics.
FEM level implements at least a queue of ready data fragments
satisfying the trigger conditions;\\
$\bullet$ SubEvB
(SubEvent Builders)
level -- data preprocessing computers grouped by detector subsystems.
SubEvB level implements at least requests of ready data fragments from the
FEM level, building of
subevents (events belonging to each detector subsystem),
queue of ready subevents, software filters for subevents rejection;\\
$\bullet$ EvB
(Event Builders)
level -- full events building computers.
EvB level implements at least requests of ready
subevents from SubEvB level, building of full events, queue
of ready full events, software filters for full events rejection;\\
$\bullet$ pool level -- data postprocessing computers.
Pool level implements at least requests of a
subset of ready events from EvB level, events conversion from a native binary
format to representation by some class of the ROOT package \cite{ROOTproc},
circle buffer of ROOT events provided
to clients for online analysis and visualization, histogramming and so on
of ROOT events, a number of these histograms provision to clients for online
analysis and visualization;\\
$\bullet$ storage level parallel to pool level -- computers, which realize
requests of ready events from EvB level and writing these events into
intermediate storage. The storage level consists of some identical computer
groups, switchable while data taking in such a way, that one group obtains
the events from EvB level when other groups transfer these data from the
intermediate into the final storage, possibly, slower than HDD.

In addition, some computer groups can be outside of the data stream:\\
$\bullet$
Slow Control group -- computers, which implement
HV and LV control and user interface, initial software
downloading into the read--out and other electronics;\\
$\bullet$
DAQ Operator group -- computers, which fulfil control and user
interface for DAQ software components;\\
$\bullet$
FEM Control group -- computers, which realize the software part
of the trigger;\\
$\bullet$
online visualization group -- clients of the pool level.

In the present paper we limit our consideration by {\itshape ngdp} key
elements only due to publishing requirements, and pend up the following
issues to the next publication: user context utilities, events
representation for the ROOT package,
control subsystem, work with CAMAC and VME hardware, simplified
``selfflow'' variants of some nodes, {\bfseries\itshape ng\_mm(4)} as
alternative to {\bfseries\itshape ng\_socket(4)}, test and debug nodes,
possible {\bfseries\itshape netgraph(4)} additionals, etc.

\vspace*{-5mm}

\section{Design and implementation}
\label{ngdp.des_impl}

\vspace*{-2mm}

\hspace*{4mm} Lets consider our requirements to the infrastructure
proposed above%
.\\
$\bullet$ Queue on the FEM level supports First Input First Output (FIFO)
discipline, which minimally allows us
to put the data packet into the end of the queue,
to get the data packet possibly of the requested type from the head
of the queue in response to the \verb|CTRL_NG_GETPACK| control packet obtaining,
to perform the queue full clear in response to the \verb|CTRL_NG_CLEAR|
 control packet or \verb|clear| control message obtaining.
This queue should be implemented by the corresponding
{\bfseries\itshape netgraph(4)} node type.
This node type provides a server functionality for the downstream (SubEvB) level
from which it obtains \verb|CTRL_NG_GETPACK| and \verb|CTRL_NG_CLEAR| control
packets (see also Table~\ref{design.ctrl_pack}), and responds to
\verb|CTRL_NG_GETPACK| by the data packet if
it is possible or -- by \verb|ANSW_NG_GETPACK| answer packet if it is not.
This node type interacts with FEM--controller by interface unspecified
here, which should, however, allow to obtain information in some pieces
to be encapsulated into the data packets, which could be put into the
queue end%
.\\
$\bullet$ Queue on the SubEvB level supports the discipline, which allows
at least as follows:
to put the data packet into the queue end;
to get the data packet (possibly of the defined type) from the queue head
(in response to \verb|CTRL_NG_GETPACK| control packet obtaining);
to get an arbitrary data packet (possibly of the defined type) from the queue
by its number (in response to \verb|CTRL_NG_GETNTHPACK| control packet
obtaining);
to perform the queue full clear (in response to \verb|CTRL_NG_CLEAR|
control packet or \verb|clear| control message obtaining).
The corresponding node type
provides a server functionality for the downstream (EvB) level,
from which it obtains \verb|CTRL_NG_GETPACK|, \verb|CTRL_NG_GETNTHPACK| and
\verb|CTRL_NG_CLEAR| control packets and responds to the former two of them
by the data packet if it is possible or -- by
\verb|ANSW_NG_GETPACK| and \verb|ANSW_NG_GETNTHPACK| answer packets.
At the same time SubEvB level functions as a client\footnote{
This is an essential feature of the proposed design -- each intermediate level
behaves as a server for the downstream level and as a client for the upstream level.
This approach simplifies algorithms of inter--level interactions, which will
be reduced to the ones only between neighbour levels.
} relatively to the upstream
(FEM) level by sending the \verb|CTRL_NG_GETPACK| and \verb|CTRL_NG_CLEAR|
control packets.\\
$\bullet$ Queue on the EvB level supports the discipline, which allows at
least as follows:
to put the data packet into the queue end;
to get the data packet (possibly of the defined type) from the queue head
 (in response to \verb|CTRL_NG_GETPACK| control packet obtaining);
to get one of each N$^{th}$ data packets (possibly of the defined type)
 without removing it from the queue (in response to \verb|CTRL_NG_COPY1OFN|
 control packet obtaining);
to perform the queue full clear (in response to \verb|CTRL_NG_CLEAR|
 control packet or \verb|clear| control message obtaining).
The corresponding node type
provides a server functionality for the downstream (pool/storage)
level, from which it obtains \verb|CTRL_NG_COPY1OFN| / \verb|CTRL_NG_GETPACK|,
\verb|CTRL_NG_CLEAR| control packets and responds to the former ones
by the data packet if it is possible or -- by \verb|ANSW_NG_COPY1OFN|
/ \verb|ANSW_NG_GETPACK| answer packets.
At the same time EvB level operates as a client relatively to the upstream
(SubEvB) level by sending the \verb|CTRL_NG_GETPACK|,
\verb|CTRL_NG_GETNTHPACK| and \verb|CTRL_NG_CLEAR| control packets.\\
$\bullet$ The pool level behaves as a client relatively to the upstream
(EvB) level by sending the \verb|CTRL_NG_COPY1OFN| control packets.
At the same time the pool level provides a server functionality for computers of
the online visualization group. This server converts each data packet into ROOT
representation of the full event (lets name it \verb|class Event|) by means of
a special constructor (or member function) of such class. After that the pool
server can:\\
 -- maintain the circle buffer of such \verb|Event|s and provide each
\verb|Event| in the form of ROOT \verb|TMessage| class instance by the
client request, or\\
 -- send each \verb|Event| as soon as possible (without bufferization)
in the form of \verb|TMessage| to each currently connected visualization
client, or to discard the corresponding data packet, if such clients are
absent, or\\
 -- fill some ROOT histogram(s) with each \verb|Event| data or collect some
statistics in another way, discard \verb|Event| itself and provide only
statistics in the form of \verb|TMessage| by the client request.

\begin{table}[htb]
\caption{Realistic queue disciplines for different data processing levels.}

\vspace*{3mm}

\begin{tabular}{|p{100mm}|c|c|c|}
\hline
Functionality, control/answer packet type and body 
              & \multicolumn{3}{|c|}{Supported on level:} \\
\cline{2-4}
contents, letter for legend
              & FEM & SubEvB & EvB \\
\hline
Get the packet from the queue head: \verb|CTRL_NG_GETPACK| with zero
body (``\verb|n|'')
  & $+$ & $+$ & $+$ \\
Answer for the above: \verb|ANSW_NG_GETPACK| with
\verb|uint16_t| error code(s) (\verb|EMPTY| only) in the body
  & $+$ & $+$ & $+$ \\
\hline
Get the packet of the defined type from the queue head: \verb|CTRL_NG_GETPACK|
with \verb|uint16_t| packet type in the body (``\verb|N|'')
  & $+$ & $+$ & $+$ \\
Answer for the above: \verb|ANSW_NG_GETPACK|
with \verb|uint16_t| error code(s) (\verb|EMPTY|, \verb|NUMNOTFOUND|,
\verb|TYPENOTFOUND|) in the body
  & $+$ & $+$ & $+$ \\
\hline
Get an arbitrary packet of the defined type from the queue by its number:
\verb|CTRL_NG_GETNTHPACK| with \verb|uint32_t| packet number and
\verb|uint16_t| packet type in the body (``\verb|G|'')
  & $-$ & $+$ & $-$ \\
Answer for the above:
\verb|ANSW_NG_GETNTHPACK| with \verb|uint16_t| error code(s)
(\verb|TYPENOTFOUND|, \verb|NUMNOTFOUND|, \verb|NUMNOTALREADY|) in the body
  & $-$ & $+$ & $-$ \\
\hline
Get one of each N$^{th}$ packets of the defined type without removing it
from the queue:
\verb|CTRL_NG_COPY1OFN| with \verb|uint16_t| N (period) and \verb|uint16_t|
packet type in the body (``\verb|O|'')
  & $-$ & $-$ & $+$ \\
Answer for the above: \verb|ANSW_NG_COPY1OFN| with \verb|uint16_t| error
code(s) (\verb|EMPTY|, \verb|NUMNOTFOUND|, \verb|TYPENOTFOUND|) in the body
  & $-$ & $-$ & $+$ \\
\hline
\end{tabular}
\label{design.ctrl_pack}

\vspace*{-2mm}

\end{table}

\begin{figure}[htb]

\vspace*{-5mm}

\epsfig{width=149mm
,file=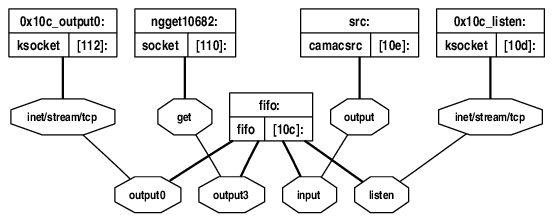
}

\vspace*{-5mm}

\caption{
A typical member of the CAMAC FEM level is realized by
the {\itshape ngdp} graph.
}
\centerline{Rectangles are nodes with: name (up), type (left), ID (right);}
\centerline{Octagons are hooks named within.}
\centerline{{\ttfamily ng\_fifo} has two output streams -- remote through
{\ttfamily accept()}ing {\ttfamily ng\_ksocket}}
\centerline{and local through {\ttfamily ng\_socket}, -- as well as
{\ttfamily listen()}ing {\ttfamily ng\_ksocket}.}
\label{pict.FEM}

\vspace*{-2mm}

\end{figure}

\begin{sidewaysfigure*}[p]
\epsfig{
width=219mm
,file=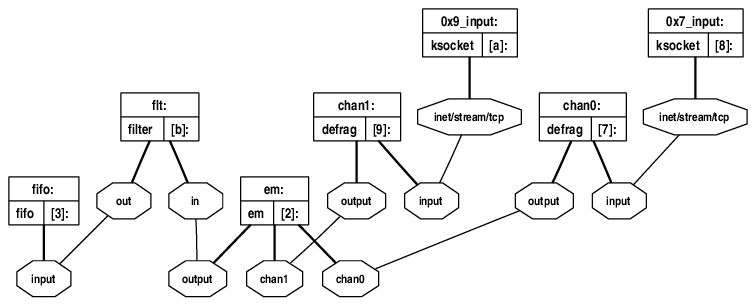
}

\caption{
A typical member of the SubEvB/EvB level is implemented by
the {\itshape ngdp} graph.
} 
\centerline{Legend the same as for Fig.~\ref{pict.FEM}.}
\centerline{{\ttfamily ng\_em} has two input channels ({\ttfamily chan0} and
{\ttfamily chan1})}
\centerline{from two data sources (for example, two crate controllers) on
FEM level (see Fig.~\ref{pict.FEM}).}
\centerline{To simplify the picture, the {\ttfamily ng\_fifo} outputs are not
shown (see Fig.~\ref{pict.FEM} for the typical ones).}
\centerline{Schematically the data packet flows in the Figure can be
represented as follows:}
\centerline{
$\begin{array}{l}
\mbox{(from} -- \to \mbox{\ttfamily ng\_ksocket} \to \mbox{\ttfamily ng\_defrag} \searrow \\
\mbox{FEM/SubEvB}
 \\
\mbox{level)} -- \to \mbox{\ttfamily ng\_ksocket} \to \mbox{\ttfamily ng\_defrag} \nearrow \\
\end{array}\mbox{\ttfamily ng\_em} \to \mbox{\ttfamily ng\_filter} \to \mbox{\ttfamily ng\_fifo}%
\begin{array}{l}
\nearrow \mbox{\ttfamily ng\_ksocket} - \to\mbox{(to} \\
~~~~~~~~~~\vdots~~~~~~~~~~~~~~\mbox{EvB/pool} \\
\searrow \mbox{\ttfamily ng\_ksocket} - \to\mbox{level)} \\
\end{array}$
}
\label{pict.SubEvB}
\end{sidewaysfigure*}

As we can see, at least three levels can contain
the same node type with slightly variated (by compiled--in or runtime
configuration) functionality, lets name it as
{\bfseries\itshape ng\_fifo(4)} (see
section~\ref{ngdp.des_impl.transp.fifo}). For example, CAMAC FEM level can
be implemented as pictured in Fig.~\ref{pict.FEM}%
: \verb|ng_camacsrc|~$\to$~\verb|ng_fifo|.
At the same time, SubEvB and EvB levels perform building of (sub)events, their
functionality can be implemented by the same {\bfseries\itshape ng\_em(4)}
(after {\itshape qdpb}'s \underline{e}vent \underline{m}erger%
) node type (see section~\ref{ngdp.des_impl.transp.em}) with configurated
requests behaviour and (sub)event building rules.
Optionally, SubEvB and EvB levels
can contain software filters for (sub)event
rejection, which reasonably could be implemented by the same
{\bfseries\itshape ng\_filter(4)} node type (see
section~\ref{ngdp.des_impl.dp.filter}) with configurated rejection rules.
So, the typical level layout (see Fig.~\ref{pict.SubEvB}) can look approximately
the following way:
\verb|ng_em|~$\to [$~\verb|ng_filter|~$\to [ \ldots ]]$~\verb|ng_fifo|%
.

\verb|ng_em| launches {\bfseries\itshape ng\_defrag(4)} (see
section~\ref{ngdp.des_impl.transp.defrag}) nodes on each
configured input channel, while \verb|ng_defrag|
launches client \verb|ng_ksocket|, which \verb|connect()|s to server
\verb|ng_ksocket| of the upstream level. After that \verb|ng_em| launches
{\bfseries\itshape kthread(9)} to send the data requests (in the control
packet
form) to the upstream level according to the configured requests mode, and
to proceed (sub)events merging in accordance with the configured building rules.

\begin{sidewaysfigure*}[p]
\epsfig{
width=219mm
,file=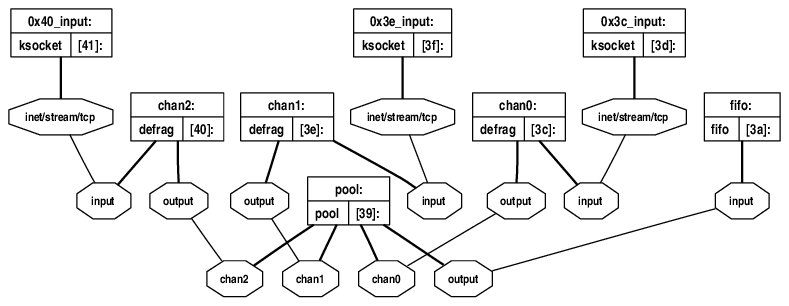
}

\caption{
A typical member of the pool level is implemented by the {\itshape ngdp} graph.
} 
\centerline{Legend the same as for Fig.~\ref{pict.FEM}.}
\centerline{{\ttfamily ng\_pool} has three input channels ({\ttfamily chan0},
{\ttfamily chan1}, {\ttfamily chan2})}
\centerline{from three data sources on EvB level
(see Fig.~\ref{pict.SubEvB}).}
\centerline{To simplify the picture, the {\ttfamily ng\_fifo} outputs are not
shown (see Fig.~\ref{pict.FEM} for the typical ones).}
\centerline{Schematically the data packet flows in the Figure can be
represented as follows:}
\centerline{
$\begin{array}{l}
\mbox{(from} -- \to \mbox{\ttfamily ng\_ksocket} \to \mbox{\ttfamily ng\_defrag} \searrow \\
\mbox{EvB} -- \to \mbox{\ttfamily ng\_ksocket} \to \mbox{\ttfamily ng\_defrag} \to \\
\mbox{level)} -- \to \mbox{\ttfamily ng\_ksocket} \to \mbox{\ttfamily ng\_defrag} \nearrow \\
\end{array}\mbox{\ttfamily ng\_pool} \to \mbox{\ttfamily ng\_fifo[s]}%
\begin{array}{l}
\nearrow \mbox{\ttfamily ng\_ksocket} - \to\mbox{(to} \\
\to ~~\mbox{\ttfamily ng\_mm} -- \to~~\mbox{{\ttfamily r2h},} \\
\searrow \mbox{\ttfamily ng\_socket} - \to~\mbox{etc.)} \\
\end{array}$
}
\label{pict.pool}
\end{sidewaysfigure*}

\verb|ng_fifo| launches server
(\verb|listen()|ing) \verb|ng_ksocket| and handles \verb|accept()|ing
\verb|ng_ksocket|(s) as needed to serve requests from the downstream level.

The pool level client can be \verb|ng_em|
in some specialized
mode (see also
section~\ref{ngdp.des_impl.transp.em}), or some separated multiplexer node
{\bfseries\itshape ng\_pool(4)} (see section~\ref{ngdp.des_impl.dp.pool}).

The pool level filter can be a node {\bfseries\itshape ng\_filter(4)} with
assistance of the user context process\footnote{
Because it is very problematic to link into kernel a C++ code in general
and ROOT classes with their dictionaries in particular.
} {\bfseries\itshape b2r(1)} (see also
section~\ref{ngdp.des_impl.dp}), or only this process.
Anyway this filter should produce ROOT \verb|Event| class instance for
each full event data packet obtained, and convert each \verb|Event|
into the so called sequential (or serialized) form using the corresponding
\verb|Streamer()| function(s).
Technically speaking, {\bfseries\itshape b2r(1)}
should use the ROOT \verb|TBufferFile| class instance to do so.
After that the sequential form of \verb|Event| has length
\verb|fBufCur|$ - $\verb|fBuffer| returned by \verb|TBufferFile::Length()|
function, should be read at \verb|TBufferFile::fBuffer| location,
prepended by packet header, and injected into netgraph again.

So, the pool level server can be a usual \verb|ng_fifo| node which
works with serialized \verb|Event|s as with usual data packets, while the
typical level layout (see Fig.~\ref{pict.pool}) can approximately look as
follows:
\verb|ng_pool|~$\to [$~\verb|ng_filter|~$\to [ \ldots ]]$~\verb|ng_fifo|%
.

Of course, we should note two additional crossings of context boundaries in
this scheme: from the kernel to the user context and back again, which can be
impractical due to too high CPU and memory consuming.

\vspace*{-5mm}

\subsection{{\itshape qdpb} inspired entities and imported elements}
\label{ngdp.des_impl.qdpb}

\vspace*{-2mm}

\hspace*{4mm} Some {\itshape ngdp} ideas and entities (see also
\cite{IsupKhNU09lat}) are inspired by the ones previously designed for the
{\itshape qdpb} \cite{IsupJINRC01-116}.
We import also the packet implementation and packet
type support from {\itshape qdpb} and redesign them in some aspects.
{\itshape qdpb}'s {\bfseries\itshape writer(1)} utility for the packet stream
writing into a regular file(s) on HDD can be used ``as is'' -- if
it is recompiled to be aware of
such changes.
 Lets note that in principle any user context
utilities previously implemented for {\itshape qdpb},
are still usable under {\itshape ngdp}, too, until
they satisfy the same condition.\\[-8mm]

\vspace*{-3mm}

\subsection{Transport subsystem}
\label{ngdp.des_impl.transp}

\vspace*{-2mm}

\hspace*{4mm} As it has been
experimentally checked, a datagram size large enough
for (local) atomic transfer through {\bfseries\itshape netgraph(4)} system
can be tuned easily. However, due to TCP/IP\footnote{
We can't use UDP/IP for many reasons, the most important of which is the
following: UDP/IP does not support datagram fragmentation while the atomic
datagram size is limited by IP packet size (64~kbytes) which generally is
too small for our purposes.
} and Ethernet\footnote{
Ethernet has a standard frame length {\ttfamily mtu}~$ = 1500$~bytes.
} network nature the
sender side unavoidably
fragments our packets, so, we should reassemble (defragment) them after
{\bfseries\itshape ng\_ksocket(4)} on the receiver side. For this
defragmentation it is enough to have
information from the packet header. Generally speaking, we have the
following options to implement a packet defragmenter:\\[-6mm]
\begin{enumerate}
\item
to compile/link the same reassembling code in many places
(practically in each of node types, discussed in
sections~ \ref{ngdp.des_impl.transp}, \ref{ngdp.des_impl.dp});\\[-6mm]
\item
to provide special KLD module with kernel--wide
implementation of the reassembling code for the nodes mentioned above;\\[-8mm]
\item
to provide special node type {\bfseries\itshape ng\_defrag(4)}.\\[-8mm]
\end{enumerate}
The latter option is the most straightforward and in a modular
{\bfseries\itshape netgraph(4)} style, it does not waste memory by the
duplicated code,
and introduces neither additional defragmenter interface nor KLD dependencies.

\vspace*{-5mm}

\subsubsection{{\bfseries \itshape ng\_defrag(4)} node}
\label{ngdp.des_impl.transp.defrag}

\vspace*{-2mm}

\hspace*{4mm} According to one of the packet defragmenter implementation
options (see 
above) we implement the first version of the packet defragmenter code using
the node of the type, which
obtains data through \verb|input| hook;
accounts their size (\verb|octets|) and number of data messages
(\verb|frames|);
defragments they into packets;
accounts a size (\verb|bytes|) and number of
resulting packets (\verb|packets|) as well as reassemble failures
(\verb|fails|) and bytes rejected during failures (\verb|rejbytes|);
stores completed packets and fragment of the last one into the circle buffer;
synchronously sends
the completed packets through the optional hook \verb|output| (if exists) or
discards them. In case of the output counterpart slower than the input one,
the node drops the packet(s) and accounts the number of
drop(s) occured (\verb|droppacks|).
This node understands the generic set
of control messages and the following specific control messages as well:\\
\verb|getclrstats| -- returns the current statistics (values of \verb|octets|,
  \verb|frames|, \verb|bytes|, \verb|packets|, \verb|fails|,
  \verb|rejbytes| and \verb|droppacks|) and
  clears it;\\
\verb|getstats| / \verb|clrstats| -- returns / clears the current
  statistics (the same values);\\
\verb|flush| -- tries to send all the packets not sent yet
  from the circle buffer.

The node supports only one hook named \verb|input| and only one hook named
\verb|output| simultaneously, and performs shutdown after all the hooks are
detached.
The node is transparent in the counterstream direction --
the data arrived through the \verb|output| hook are sent ``as is'' through
the \verb|input| hook.

Later we have improved this node to launch client
\verb|ng_ksocket|, to \verb|connect()| it to server
\verb|ng_ksocket| of the upstream level, and to attach it
to the \verb|input| hook. This {\bfseries\itshape ng\_defrag(4)} node
understands two additional
control messages:\\
\verb|connect <struct sockaddr addr>| -- supplies IP address/port (in the
same format
as understood by {\bfseries\itshape ng\_ksocket(4)} node) of the server to
\verb|connect()| with;\\
\verb|needchknum <int8_t flag>| -- (re)sets \verb|flag|, which means to apply
\verb|checknum()|
for each reassembled packet if the \verb|flag| is nonzero,
otherwise it is not applied.

\vspace*{-5mm}

\subsubsection{{\bfseries \itshape ng\_fifo(4)} prototype}
\label{ngdp.des_impl.transp.fifo}

\vspace*{-2mm}

\hspace*{4mm} In order to implement node
{\bfseries\itshape ng\_fifo(4)} with buffer disciplines, described in
section~\ref{ngdp.des_impl}, as a first step we implement
some prototype, which is able to:\\
$\bullet$ spawn \verb|listen()|ing \verb|ng_ksocket| at startup;\\
$\bullet$ spawn \verb|accept()|ing \verb|ng_ksocket|(s) at each connection
 request (up to the configured maximum) from the known host(s)/port(s), and/or
 accept the hook connect from the local netgraph \verb|ng_socket|(s);\\
$\bullet$ emit each data packet obtained on the \verb|input| hook (or
 internally generated
 if such hook is absent) in response to request\footnote{
Lets call such output policy as {\ttfamily LAZY} in contrast
with {\ttfamily ASAP} (As--Soon--As--Possible)%
.
 } (in the form of the control packet)
 obtaining through only the same \verb|accept()|ing \verb|ng_ksocket|
 or local \verb|ng_socket|;\\
$\bullet$ close \verb|accept()|ing \verb|ng_ksocket| at EOF notification
 obtaining or connection losing.

Such functionality does not require
kernel thread usage,
however can neither respawn \verb|listen()|ing \verb|ng_ksocket| in case of
its shutdown due to some external or accidental reasons,
nor handle nontrivial internal errors during \verb|listen()|ing
\verb|ng_ksocket| initialization. The reason is impossibility of using
(at least with macroscopic timeouts) {\bfseries\itshape msleep(9)} in the
context, where
{\bfseries\itshape netgraph(4)} code is executed (usually one of the
{\bfseries\itshape swi(9)} software interrupt threads). So, this additional
error handling requires {\bfseries\itshape kthread(9)} usage and can
be implemented without ideological or technical problems.

The prototype supports universal queue discipline
``\verb|nNGO|'' (see Table~\ref{design.ctrl_pack}), which is suitable for FEM,
SubEvB, EvB and pool level bufferization simultaneously and provides all
queue access kinds, which required to support
\verb|CTRL_NG_CLEAR|
 (with and without \verb|ptype| argument),
\verb|CTRL_NG_GETPACK|
(with and without \verb|ptype| argument),
\verb|CTRL_NG_COPY1OFN(period, ptype)|, ~and
~\verb|CTRL_NG_GETNTHPACK(pnum, ptype)|
control packet types.
We implement this universal queue first of all as the user context model
{\itshape tbuf\_nNGO.c}
and debug such model strongly, to be sure that this
implementation is working now.

The \verb|ANSW_*| packet bodies contain one of the
following error codes as \verb|uint16_t| value (see
Table~\ref{design.ctrl_pack}) to provide more information to client nodes
({\bfseries\itshape ng\_em(4)}, {\bfseries\itshape ng\_pool(4)}, etc.)
to make up a decision:\\
\verb|EMPTY| (``\verb|n|'', ``\verb|N|'' and ``\verb|O|'' buffer
operations\footnote{
Here only realistic (with {\ttfamily ptype} argument) buffer operations
``{\ttfamily nNGO}'' (see Table~\ref{design.ctrl_pack}) are mentioned.
}) -- buffer is empty now;\\
\verb|TYPENOTFOUND| (all buffer operations) -- requested packet type not yet
obtained;\\
\verb|NUMNOTFOUND| (all buffer operations) -- requested packet number not yet
obtained;\\
\verb|NUMNOTALREADY| (``\verb|G|'' buffer operation) -- requested packet number
already dropped from the buffer.

To simplify \verb|mkpeer|ing in some situations, the
{\bfseries\itshape ng\_fifo(4)} node supports the \verb|creat| hook, which can be
removed after the \verb|input| or \verb|listen| hook appears, however the
\verb|input| hook can be used for \verb|mkpeer|ing, too, if this is convenient.
The prototype understands the generic set of control messages as well as the
following specific
ones:\\
\verb|start|/\verb|stop| -- allows/denies getting packets from the queue;\\
\verb|lstnaddr| -- sets
  IP address and port to \verb|bind()| our
  \verb|listen()|ing \verb|ng_ksocket|;\\
\verb|addaddr|/\verb|deladdr| -- adds/deletes network IP address and port from
  which connection requests should be \verb|accept()|ed by our
  \verb|ng_ksocket|;\\
\verb|getclrstats| -- returns the current statistics (numbers of
  \verb|packets_out|,
  \verb|bytes_out| and \verb|fails|, \verb|elapsed| and \verb|pure| times) and
  clears it;\\
\verb|getstats| / \verb|clrstats| -- returns / clears the current statistics
  (the same values).

\vspace*{-2mm}

\subsubsection{{\bfseries \itshape ng\_em(4)} prototype}
\label{ngdp.des_impl.transp.em}

\vspace*{-2mm}

\hspace*{4mm} In order to implement a node
with {\bfseries\itshape ng\_em(4)} functionality, described in
section~\ref{ngdp.des_impl}, as a first step
we implement some prototype, which is able to:\\
$\bullet$ launch \verb|ng_defrag| node at each configured input channel, which
launches client \verb|ng_ksocket| node to \verb|connect()| to the upstream
server corresponding to the channel;\\
$\bullet$ send requests in the form of control packets according to the
working mode (one of ``SubEvBt'' or ``EvBt'') that has been configured;\\
$\bullet$ merge packets obtained on the input channels according to the
merging rules which have been configured.

Generally (with some simplifications) speaking, in the SubEvBt working mode
the prototype makes one loop over the configured merging rules (and corresponding
requests) array and launches the kernel thread (see {\bfseries\itshape kthread(9)})
for each configured index, so each thread serves only its ``own'' request.
Each thread emits \verb|CTRL_NG_GETPACK(ptype)| control packets (see also
Table~\ref{design.ctrl_pack}) through the hooks of the involved
input channels. After that each thread waits for responses in the form of the
data packets (always means positive response) and/or answer packets (always
means negative response) up to obtaining all the required packets
or corresponding (regular) timeout expiration. If the answer packet(s) is
obtained, the thread analyses the error code(s) and
either cleans the input channel storages and sends the full request again, or
repeats request(s) in the failed input channel(s) (after
either the same or increased regular timeout).
If some input channel(s) does not respond at all before regular timeout
expiration, the thread analyses the state of the responded channels and
either repeats request(s) in the failed input channel(s), or cleans the input
channel storages and sends the full request again. The regular timeout can be
increased up to the limit only. If all the required data packets are obtained,
the prototype merges them into a resulting packet and sends it to the
\verb|output| hook (if
any). After that the thread sets a regular timeout to the nominal value,
sends the full request again, and so on.

In the EvBt working mode the prototype makes one loop over the configured
requests array and launches the kernel thread to serve each configured index,
too. Each request has the so called trigger input channel and is handled in
two phases. In the first (Trig) phase each thread emits \verb|CTRL_NG_GETPACK(ptype)|
control packet (see also Table~\ref{design.ctrl_pack}) through the hook of the
trigger input channel and waits for a positive or negative response
up to obtaining one or corresponding (trigger) timeout expiration.
If the answer packet is obtained, the thread analyses the error code and repeats
the request after either the same or increased trigger timeout.
If the trigger input channel does not respond at all before the trigger
timeout expiration, the thread repeats the request and waits during the
increased trigger timeout.
The trigger timeout can be increased up to the limit only, too.
If the data packet from the trigger server is successfully obtained, the
prototype extracts $N$ number\footnote{
$T$ type is also extracted and
checked against
the resulting type of the corresponding merging rule.
} from its body and goes to the second phase, which for each request index is
handled by the same thread as the first phase.
In the second (afterTrig) phase the thread emits \verb|CTRL_NG_GETNTHPACK(|$N$\verb|, ptype)|
control packets (see also Table~\ref{design.ctrl_pack}) through the hooks
of the involved input (other than trigger) channels using $N$ mentioned above
and waits for positive and/or negative responses up to obtaining all the
required packets
or regular timeout expiration.
After that the algorithm behaves
as it is described above for the SubEvBt mode.

Note that all these working modes require servers (\verb|ng_fifo| nodes)
with the support of the corresponding queue disciplines (as described in
section~\ref{ngdp.des_impl}).

Duties between the kernel thread(s) and synchronous parts of the prototype are
separated as follows: each
\verb|rcvdata()| execution processes single packet, possibly calls
\verb|evmerge()| or \verb|evclean()|, and
either sets a special flag \verb|kth_need|
and wakes the thread up, or not. So, the thread can be waken up by the external
event ({\itshape ngdp} packet or {\bfseries\itshape netgraph(4)} control
message arriving, etc.) or after the timeout expiration. In the first case the
thread performs some actions according to \verb|kth_need| flag value, and
sets the transition state flag \verb|kth2state|. In the second case it
performs some actions according to the \verb|kth2state| flag value and sets
it again. In the both cases the thread possibly calls \verb|sendreq()| and
\verb|evmerge()| or \verb|evclean()|, and finally goes to
{\bfseries\itshape msleep(9)} with the corresponding timeout again.

We implement such nontrivial {\bfseries\itshape ng\_em(4)}'s algorithm as
single source able to be compiled for both the
kernel context using {\bfseries\itshape kthread(9)} -- for production purposes,
and the user context
using {\bfseries\itshape pthread(3)} -- for debug purposes.

The scheme of the packet requests assumes that only the packets with equal
numbers can be merged.
Later we generalize this approach and introduce id mark -- some entity
from the packet header to be compared (really subtracted)
for each two candidates for being merged. Up to \verb|MAX_ID| of these
id marks can be configured. The first added id mark is compared first.
Comparison functions of all the configured ids should return zero to permit
merging. In the current packet header implementation it is reasonable to choose
the following header fields as id marks:\\
the packet number -- id mark named \verb|"num"|, function \verb|cmp_num()|
returns zero for equal packet numbers;\\
the time stamp -- id mark named \verb|"tv"|, function \verb|cmp_tv()| returns
zero if time stamps are closer than the supplied function argument \verb|arg|
(in mksec).

The \verb|"num"| id mark is added at startup (in the node constructor) to
provide the expected node behaviour by default.

To simplify \verb|mkpeer|ing in some situations, the
{\bfseries\itshape ng\_em(4)} node supports the \verb|creat| hook, which can be
removed after \verb|input<N>| or \verb|output| hook appearing, however the
\verb|output| hook can be used for \verb|mkpeer|ing, too, if this is convenient.
The prototype understands the generic set of control messages as well as the
following specific
ones:\\
\verb|getclrstats <char *inchan>| -- returns the current
  statistics (values of \verb|packets_in|, \verb|bytes_in| and \verb|reqs|)
  and clears it for the input
  channel named \verb|<inchan>|;\\
\verb|getstats <char *inchan>| / \verb|clrstats <char *inchan>| --
  returns / clears the current statistics
  (the same values) for \verb|<inchan>|;\\
\verb|getclrostats| -- returns the current statistics (values of
  \verb|packets_out|,
  \verb|bytes_out| and \verb|fails|) and
  clears it for the \verb|output| hook;\\
\verb|getostats| / \verb|clrostats| -- returns / clears the current statistics
  (the same values) for the \verb|output| hook;\\
\verb|flush| -- marks buffers of all the input channels as empty;\\
\verb|inchan <struct ng_em_cfgentry>| -- adds configuration
  entry to introduce new input channel according to supplied
  \verb|<ng_em_cfgentry>| members: name of the input channel \verb|char *name|,
  trigger bit \verb|int8_t trig| for it (for SubEvBt mode means
  nothing), IP address \verb|struct sockaddr addr| to connect, number of the
  request
  entry \verb|int8_t idx|, request entry configuration \verb|struct emtbl|
  (see below), -- and \verb|mkpeer|s needed \verb|ng_defrag| nodes;\\
\verb|getinchan <char *inchan>| -- returns configuration of the \verb|<inchan>|
  input channel;\\
\verb|addcfg <struct ng_em_tblentry>| -- adds the request entry to the
  already existing
  input channel according to the supplied \verb|<ng_em_tblentry>| members:
  name of the
  input channel \verb|char *name|, trigger bit \verb|int8_t trig| for it (for
  SubEvBt mode means nothing), number of the request entry \verb|int8_t idx|,
  request entry configuration \verb|struct emtbl tbl| with the
  \verb|uint16_t in_type|, \verb|uint16_t out_type|, \verb|u_char order|,
  and \verb|u_char number| mandatory members;\\
\verb|delcfg <struct ng_em_tblentry>| -- deletes the already existing request
  entry of the input channel \verb|<name>| by \verb|<idx>| or (for
  \verb|<idx>| equals to \verb|-1|) by \verb|tbl.in_type|;\\
\verb|getreq <int8_t idx>| -- returns configuration of the full request
  with number \verb|<idx>|;\\
\verb|delreq <int8_t idx>| -- deletes configuration of the full request
  with number \verb|<idx>| (equivalent to do \verb|delcfg| for each input
  channel involved into such request);\\
\verb|connect <char *mode>| -- checks the already supplied input channels and
  request entries configuration to operate in \verb|<mode>| (valid are
  \verb|"SubEvBt"| or \verb|"EvBt"|), removes the
  unused input channels (if any) and connects the not yet connected
  \verb|ng_defrag|(s) to servers according to the current configuration;\\
\verb|start <int64_t num_of_reqs>| -- starts request sending by thread(s)
  up to the \verb|<num_of_reqs>| requests will be issued;\\
\verb|stop| -- immediately stops the request sending%
;\\
\verb|addcmp <struct ng_em_addcmp>| -- adds id mark comparison function
  described by structure
  \verb|<ng_em_addcmp>|, which supplies the function name
  \verb|char *name| and arguments array \verb|union arg arg_arr[]|;\\
\verb|delcmp <char *name>| -- deletes id mark comparison function named
  \verb|<name>|;\\
\verb|clrcmp| -- clears whole id mark(s) configuration;\\
\verb|getcmp| -- returns the full current id mark(s) configuration;\\
\verb|gettrig| -- returns the full current configuration of the trigger input
  channels (for SubEvBt mode means nothing);\\
\verb|setsubnames <struct ng_em_subnames>| -- sets naming style for
  \verb|ng_defrag| sub\-no\-des as defined by \verb|int8_t mode| structure
  member, which can be equal to the following values \verb|#define|d in
  {\itshape ng\_em.c}:\\
  \hspace*{4mm}\verb|SUBNAMES_NONE| (does not name subnodes at all),\\
  \hspace*{4mm}\verb|SUBNAMES_TYPICAL| (names by corresponding \verb|inchan|
  name -- startup default),\\
  \hspace*{4mm}\verb|SUBNAMES_UNIQUE| (uses the unique node ID in name),\\
  \hspace*{4mm}\verb|SUBNAMES_PREF| (prepends by the supplied string),\\
  \hspace*{4mm}\verb|SUBNAMES_SUFF| (appends by the supplied string),\\
  where the
  \verb|char *str| member is a prefix or suffix used by the last two modes;\\
\verb|setsubtype <char *type>| -- sets node type is welcomed to
  connect as input channel(s): default is \verb|"defrag"|, empty string
  \verb|""| means any type can be connected, node types other than \verb|"defrag"|
  are not launched automatically by \verb|inchan| control message, so it should
  be followed by the explicit \verb|connect| control message;\\
\verb|getsubtype| -- returns the current subnode type;\\
\verb|settimo <struct ng_em_settimo>| -- sets the timeout configuration according
  to the supplied \verb|<struct ng_em_settimo>| members: number of request entry
  \verb|int8_t idx|, base timeout value \verb|int32_t t_timo| (in msecs) and the
  timeout increasing limit factor \verb|int32_t t_f| for the Trig phase (in
  SubEvBt mode means nothing), the same for the afterTrig phase --
  \verb|int32_t r_timo| (in msecs) and \verb|int32_t r_f|,
  limit for the number of request failures \verb|int32_t r_max|;\\
\verb|gettimo| -- returns the current timeout configuration as
  \verb|<struct ng_em_settimo>| for each existing request entry.

{\bfseries\itshape ng\_em(4)} prototype supports only one hook named
\verb|output| simultaneously, and
is transparent in the counterstream
direction for debug purposes --
the data arrived through the \verb|output| hook are sent ``as is'' through the
hook, which corresponds to the input channel with zero number.

\vspace*{-5mm}

\subsection{Data processing subsystem}
\label{ngdp.des_impl.dp}

\vspace*{-2mm}

\hspace*{4mm} Lets consider the {\itshape ngdp} elements used for some data
transformations.\\
1. Pool of events -- has the following implementation options:\\
 $\bullet$ Possible pool level layout can be described by the following graph\\
\centerline{
$\mbox{\ttfamily ng\_pool}%
\stackrel{\textstyle\nearrow}{\vphantom{-}}\begin{array}{c}
 \mbox{\ttfamily ng\_filter} \\
 \downarrow \uparrow \\
 \mbox{\ttfamily b2r}
\end{array}\stackrel{\textstyle\searrow}{\vphantom{-}}%
\mbox{\ttfamily ng\_fifo} \ \ ,$
}
where
{\bfseries\itshape ng\_pool(4)} is {\bfseries\itshape ng\_em(4)} in
some specialized working mode, or some separately implemented node with a
very similar functionality.
This approach is impossible without assistance of the user context process(es)
{\bfseries\itshape b2r(1)}, which converts
each obtained packet into ROOT representation of the full event
(\verb|class Event|)
and serializes them using \verb|class TBufferFile|
instances. Bufferization of the serialized \verb|Event|s done by
{\bfseries\itshape ng\_fifo(4)} node%
. Additional double data copying
from the kernel to the user context and back again should be noted.\\
 $\bullet$ 
Some server (user context process%
), which obtains the data packets from a single input stream multiplexed by
{\bfseries\itshape ng\_pool(4)}\footnote{
It issues the corresponding control packets to
request data from the upstream (EvB) level.
}, converts each packet
into ROOT representation of the full event
(\verb|class Event|)%
, maintains the memory based
``pool'' of such events, and sends the events in the form of
\verb|TMessage| ROOT class instances at client requests. This pool could
be a circle queue with two possible update policies:
lazy -- by last reader, or contemporar -- by data appearing on EvB%
.\\
2. {\bfseries\itshape ng\_filter(4)} -- node provides the software filter
functionality for (sub)events rejection, possibly located between
{\bfseries\itshape ng\_em(4)} and {\bfseries\itshape ng\_fifo(4)} on SubEvB
/ EvB level.

\vspace*{-2mm}

\subsubsection{{\bfseries \itshape ng\_pool(4)} prototype}
\label{ngdp.des_impl.dp.pool}

\vspace*{-2mm}

\hspace*{4mm}
Currently we decide to implement the pool level functionality by
{\bfseries\itshape ng\_pool(4)} node
separately from the
very similar {\bfseries\itshape ng\_em(4)}. The first stage prototype is able
to:\\
$\bullet$ launch the \verb|ng_defrag| node at each configured input channel, this
node in its turn launches the client \verb|ng_ksocket| node, which
\verb|connect()|s to the upstream server corresponding to this channel;\\
$\bullet$ send requests in the form of control packets;\\
$\bullet$ transmit all packets, accepted in input
channel(s) according to the configured rules, through the \verb|output| hook.

The prototype makes one loop over the requests
array and launches the kernel thread for each configured index.
Through the hook of each involved input channel each thread emits
\verb|CTRL_NG_COPY1OFN(ptype)|
control packet (see also Table~\ref{design.ctrl_pack}) and
waits for positive or negative responses
in the form of data or answer packet until the packet is
obtained or the corresponding timeout is expired.
If the thread obtains the answer packet in some input channel, it marks such
channel to be requested again after the timeout expiration.
If some input channel does not respond in any form during the full timeout,
the thread performs the request in this channel again.
If the prototype obtains the data packet in some input channel, it sends
this packet without changes to the \verb|output| hook (if any), and emits the
request in this channel again.

Of course, servers (\verb|ng_fifo| nodes)
with the support of the corresponding queue discipline (as described in
section~\ref{ngdp.des_impl}) are required.

The algorithm described above like the one from {\bfseries\itshape ng\_em(4)}
(see section~\ref{ngdp.des_impl.transp.em})
essentially requires to use {\bfseries\itshape kthread(9)}.
Using the same approach as mentioned for {\bfseries\itshape ng\_em(4)}
we can compile a single source
for both the kernel context
and the user context%
.
After a strong debug sessions in the both contexts we are sure to have worked
out {\bfseries\itshape ng\_pool(4)} algorithm
implementation now.
The prototype understands the generic set of control messages as well as
the following specific
ones:\\
\verb|getclrstats <char *inchan>| -- returns the current statistics
  (values of \verb|data_packs|, \verb|data_bytes|, \verb|answ_packs|,
   \verb|answ_bytes|, \verb|fails|, \verb|refus| and \verb|reqs|) and
  clears it for the input channel named \verb|<inchan>|;\\
\verb|getstats <char *inchan>| / \verb|clrstats <char *inchan>| --
  returns / clears the current statistics (the same values) for
  \verb|<inchan>|;\\
\verb|getclrostats| -- returns the current statistics (values of
  \verb|packets_out|, \verb|bytes_out| and \verb|fails|) and
  clears it for the \verb|output| hook;\\
\verb|getostats| / \verb|clrostats| -- returns / clears the current
  statistics (the same values) for the \verb|output| hook;\\
\verb|addcfg <struct tbl>| / \verb|delcfg <struct tbl>| -- adds / deletes
  the request entry;\\
\verb|getconf| -- returns the full current request configuration;\\
\verb|inchan <struct ng_pool_cfgentry>| -- adds the configuration
  entry to introduce a new input channel and
  \verb|mkpeer|s needed \verb|ng_defrag| node;\\
\verb|connect| -- connects \verb|ng_defrag|(s) to servers and
  launches/terminates threads according to the current configuration of the
  input channels and request entries;\\
\verb|delinchan <char *name>| -- disconnects (if needed) and deletes the input
  channel named \verb|<name>| (node should be in the \verb|stop| state);\\
\verb|start <struct ng_pool_start>| -- starts request sending by thread(s)
  with supplied \verb|<idx>| up to the \verb|<num_of_reqs>| requests will
  be issued,
  \verb|<idx>| equals to \verb|-1| activates all configured thread(s);\\
\verb|stop <int8_t idx>| -- immediately stops the request sending by thread
  with \verb|<idx>|%
;\\
\verb|allow <disposition>| -- (re)sets allow/deny disposition of the input
  channels according to the supplied \verb|<disposition>| array of
  \verb|int8_t|: positive values mean to allow the packet
  obtaining, negative -- to deny, zero -- not to change;\\
\verb|getallow| -- returns the current allow/deny disposition for all the
  configured input channels;\\
\verb|settimo <struct ng_pool_timo>| / \verb|gettimo| -- sets / returns the
  nominal timeout (in msecs) and multiplier values.

{\bfseries\itshape ng\_pool(4)} prototype supports only one hook named
\verb|output| simultaneously.
{\bfseries\itshape ng\_pool(4)} prototype transparent in the counterstream
direction for debug purposes -- the
data arrived through the \verb|output| hook are sent ``as is'' through the
hook, corresponding to the input channel with zero number.

\vspace*{-3mm}

\subsubsection{{\bfseries \itshape ng\_filter(4)} prototype}
\label{ngdp.des_impl.dp.filter}

\vspace*{-2mm}

\hspace*{4mm} As a first step of implementation a node
with {\bfseries\itshape ng\_filter(4)} functionality described in
section~\ref{ngdp.des_impl.dp}, some prototype is released, which is able to:\\
$\bullet$ insert itself between two already connected
foreign hooks, using two own hooks, \verb|in| and \verb|out|;\\
$\bullet$ restore the situation before insertion;\\
$\bullet$ stay without any hooks to allow another insertion(s);\\
$\bullet$ connect external
filter implementation -- (pipe of) user context process(es)\footnote{
F.e., {\ttfamily ngget filter subout | b2r -O | ngput filter subin}~~.
} or (chain of) netgraph node(s) -- using two additional hooks,
\verb|subout| and \verb|subin|;\\
$\bullet$ filter nothing (dummy internal filter procedure).

For initial \verb|mkpeer|ing of {\bfseries\itshape ng\_filter(4)} a
specialized hook \verb|creat| should be used, which will be removed automatically
after successful \verb|insert|ion (or can be removed manually at any time).
The prototype supports the following specific control messages:\\
\verb|getclrstats| -- returns the current statistics
  (\verb|in_packets|, \verb|out_packets|, \verb|in_bytes|,
  \verb|out_bytes| values) and clears it for each of
  \verb|in|, \verb|subout|, \verb|subin| and \verb|out| hooks;\\
\verb|getstats| / \verb|clrstats| -- returns / clears the current statistics
  (the same values);\\
\verb|insert "<path1>/<path2>"| -- breaks the existing connection and
  connects the own \verb|in| hook to hook, represented by \verb|<path1>|, and
  the own \verb|out| hook -- to \verb|<path2>|;\\
\verb|bypass| -- removes itself and reconnects peer hooks as it was before
  the last \verb|insert|ion.

At the same time
the \verb|in| (\verb|out|)
hook can be created separately from the \verb|out| (\verb|in|) hook during
the usual \verb|mkpeer| or \verb|connect| procedures. Note, however, that
\verb|rmhook|ing of the \verb|in| (\verb|out|) hook leads to removing the
\verb|out| (\verb|in|) hook, too, without peer hooks reconnection\footnote{
Due to {\bfseries\itshape netgraph(4)}'s nature of the hook disconnection.
}, that is why {\bfseries\itshape ng\_filter(4)}'s hook removing should be
avoided. In contrast, {\bfseries\itshape ng\_filter(4)}'s shutdown sequence
performes such reconnection graciously before the node is over.

As the next step we implement some kind of ``plug--in'' mechanism which
allows us to load and unload the internal filter procedures implemented as KLD
modules during the {\bfseries\itshape ng\_filter(4)} node runtime.
Namely, the filter procedure
under the
name {\ttfamily xxx} should be by convention
in the KLD module named
{\ttfamily flt\_xxx} stored in the file {\itshape flt\_xxx.ko}~. This module
should contain a {\ttfamily void *flt\_xxx\_ptr} variable which points to
{\ttfamily struct ng\_filter\_flt flt\_xxx\_arr[]} -- container for one or more
filter function\footnote{
Prototyped as {\ttfamily int (*fltfunc\_t)(item\_p, union arg *)}~.
} pointer(s) as well as argument(s), name(s) and some flags. So after
fresh \verb|mkpeer|ing the {\bfseries\itshape ng\_filter(4)} node instance
appears without any filter procedures. After that any filter
procedure(s) can be registered by the \verb|addflt| control message (see below)
at any time. This leads to corresponding KLD module loading (if not yet) and
filter adding (if not yet) at the end of the filter procedure chain. For each
{\bfseries\itshape netgraph(4)}'s \verb|item| obtained from the \verb|in|
hook each chain member will be applied sequentially, starting from the beginning
of the chain, up to the first nonzero procedure return or chain end. In the former
case the \verb|item| is freed, in the latter it passes through the \verb|out|
(or \verb|subout|, if any) hook. Each \verb|item| from the \verb|subin| hook
(if any) passes untouched through the \verb|out| hook. Any filter procedure
can be deregistered by the \verb|delflt| control message (see below) at any
time, which leads to filter deleting from the filter procedure chain and the
corresponding KLD module unloading (if no longer referred to by anybody).

The following specific control messages were added:\\
\verb|addflt <struct ng_filter_addflt>| -- adds the filter procedure with
  supplied name
  \verb|<char *name>| as the last procedure of the filter chain and fills the
  array
  of its arguments from the supplied \verb|<union arg arg_arr[]>|;\\
\verb|delflt <char *name>| -- removes the filter procedure named \verb|<name>|
  (if any) from the
  filter chain and unloads the corresponding KLD module if no longer used
  by other instance(s) of {\bfseries\itshape ng\_filter(4)};\\
\verb|getflt| -- returns the array of \verb|<struct ng_filter_addflt>| (the full
  filter chain configuration);\\
\verb|clrflt| -- clears the whole filter chain.

As it has already been mentioned, the user context process participation in
filtering
will be unavoidable in some cases (f.e., events conversion to ROOT
representation). To allow a more efficient way for the packets to cross the
context boundaries twice -- from the kernel to the user and back again -- the
{\bfseries\itshape ng\_mm(4)} node
can be applied.
This node can be connected
to {\bfseries\itshape ng\_filter(4)} \verb|subout| and \verb|subin| hooks
by its \verb|in| and \verb|out| hooks, correspondingly. A user context process
can map the both buffers (for raw and converted
packets) into the own address space by {\bfseries\itshape mmap(2)} mechanism,
supported by {\bfseries\itshape ng\_mm(4)}'s {\itshape /dev/mmr}\verb|<N>|
and {\itshape /dev/mmc}\verb|<N>| devices. After that the process can directly
communicate\footnote{
This also eliminates possible packet fragmentation by
{\bfseries\itshape ng\_socket(4)} as well as socket I/O buffer size issues.
} with these buffers as with regular pieces of memory. Of course,
some synchronization is required and can be done by calling
{\bfseries\itshape ioctl(2)} to such devices before and after the buffer
reading and writing.

\vspace*{-3mm}

\subsection{User context utilities}
\label{ngdp.des_impl.utils}

\vspace*{-2mm}

\hspace*{4mm} To simplify the data exchange between the user and kernel context
entities of the {\itshape ngdp} system, the
{\bfseries\itshape ngget(1)} and {\bfseries\itshape ngput(1)}
utilities are implemented. A standard {\bfseries\itshape netgraph(4)}'s way
to do such exchange is to communicate
through {\bfseries\itshape ng\_socket(4)} node, which at the same time is a
socket in the specific domain. However, for speed reasons we have also
implemented
{\bfseries\itshape ng\_mm(4)} node%
, which at the same time is a UNIX device with support of the
{\bfseries\itshape mmap(2)} mechanism.
This provides us the option to read the packets from
the circle buffers allocated in the kernel directly
and write them there,
instead of flowing the packets through a number of layers
of the socket machinery. The {\bfseries\itshape ngget(1)} and
{\bfseries\itshape ngput(1)} are able to use both
{\bfseries\itshape ng\_socket(4)} and {\bfseries\itshape ng\_mm(4)} mechanisms.
The {\bfseries\itshape ngget(1)} reads the packets from the kernel graph and
writes them to the standard output.
The {\bfseries\itshape ngput(1)} reads the packets from the standard input and
writes them to the kernel graph.

\vspace*{-5mm}

\section{Conclusions}
\label{ngdp.5}

\vspace*{-3mm}

\hspace*{4mm} Using the {\bfseries\itshape netgraph(4)} system
we have demonstrated a possibility of implementing
the data transportation and processing framework {\itshape ngdp} for
the DAQ system building. The {\itshape ngdp} is as modular, lightweight
and fast as possible under an ordinary UNIX--like OS. Several kernel
context modules and user context utilities for the {\itshape ngdp} system have
been designed, implemented and debugged.

\vspace*{-5mm}


\end{document}